\documentclass[letterpaper, 10pt, conference]{ieeeconf}    
\IEEEoverridecommandlockouts                            

\usepackage{graphicx}   
\usepackage{ascmac}

\usepackage{bm}          
\usepackage{bbm}         
\usepackage{amsmath}     
\usepackage{amssymb}    
\usepackage{amsfonts}    
\usepackage{mathtools}

\usepackage{amsthm}      
\usepackage{algorithmic}
\usepackage{algorithm}   
\usepackage{color}      
\usepackage{wrapfig}     
\usepackage{url}         
\usepackage{cite}        
\usepackage{color}      
\usepackage{comment}     
\usepackage{verbatim}  
\usepackage{physics}

\theoremstyle{definition}

\newtheorem{theorem}{Theorem}          
\newtheorem{assumption}{Assumption}    
      
\newtheorem{lemma}{Lemma}

\newtheorem{remark}{Remark}

\usepackage{hyperref}
\usepackage{xurl}

\title{\LARGE \bf
Bounding Transient Moments for a Class of Stochastic Reaction Networks
Using Kolmogorov's Backward Equation
}

\author{Takeyuki Iwasaki and Yutaka Hori
\thanks{This work was supported in part by JSPS KAKENHI Grant Numbers JP24K00911, JP23H00506, JP24KK0267, JP26H00483, and JP26K00968, and by the JST-Mirai Program under Grant Number JPMJMI25H2.}
\thanks{T. Iwasaki and Y. Hori are with the Department
of Applied Physics and Physico-Informatics, Keio University, Kanagawa 223-8522 Japan. Correspondence should be addressed to Y. Hori. {\tt yhori@keio.jp}}
}

\begin{document}

\maketitle
\thispagestyle{empty}
\pagestyle{empty}

\begin{abstract}
Stochastic chemical reaction networks (SRNs) in cellular systems are commonly modeled as continuous-time Markov chains (CTMCs) describing the dynamics of molecular copy numbers.
The exact evaluation of transient copy number statistics is, however, often hindered by a non-closed hierarchy of moment equations.
In this paper, we propose a method for computing theoretically guaranteed upper and lower bounds on transient moments based on the Kolmogorov's backward equation, which provides a dual representation of the CME, the governing equation for the probability distribution of the CTMC. This dual formulation avoids the moment closure problem by shifting the source of infinite dimensionality to the dependence on the initial state.
We show that, this dual formulation, combined with the monotonicity of the CTMC generator, leads to a finite-dimensional linear time-invariant system that provides bounds on transient moments. 
The resulting system enables efficient evaluation of moment bounds across multiple initial conditions by simple inner-product operations  without recomputing the bounding system.
Further, for certain classes of SRNs, the bounding ODEs admit explicit construction from the reaction model, providing a systematic and constructive framework for computing provable bounds.

\end{abstract}

\section{Introduction}\label{sec:introduction}

Stochastic chemical reaction networks (SRNs) are widely used to analyze and design the intracellular biochemical processes, where chemical reactions occur in an inherently stochastic manner due to the low copy numbers of reactant molecular species \cite{khammash2022cybergenetics,briat2023noise}. 
In this regime, the stochastic dynamics of the molecular copy numbers are modeled as a continuous-time Markov chain (CTMC) \cite{anderson2011continuous} on an integer lattice representing the copy numbers of each molecular species.
This stochastic process is governed by the chemical master equation (CME), which takes the form of the Kolmogorov's forward equation \cite{gillespie1992rigorous}.
A major challenge in analyzing the CME, however, is its infinite-dimensional nature arising from the unbounded state space due to the unboundedness of molecular copy numbers, 
 making analytical solutions difficult to obtain.
This has led to significant research efforts to develop approximate computational algorithms \cite{gillespie2000chemical, gillespie1976general, munsky2006finite, sukys2022approximating}, including Monte Carlo simulation methods \cite{gillespie1976general} and state-truncation-based approaches such as the finite state projection (FSP) \cite{munsky2006finite}.

On the other hand, moment-based approaches describe the evolution of statistical quantities such as moments of molecular copy numbers through moment equations derived from the CME, enabling intuitive analysis of stochastic behaviors.
However, in many practical cases, the resulting moment equations again form an infinite hierarchy, as low-order moments depend on higher-order ones, resulting in an infinite dimensional ODE as with the CME.
Hence, this approach has also motivated significant efforts to develop approximate computational algorithms. 
Among them, moment closure is a widely used method to \textit{close} the equation by approximating the high-order moments using low-order ones \cite{naasell2003extension,hespanha2008moment,singh2010approximate,naghnaeian2017robust}.
However, a main drawback is the lack of rigorous error guarantees. 
Recently developed optimization-based approaches address this issue by providing guaranteed bounds on moments \cite{ghusinga2017exact,sakurai2017convex, sakurai2018optimization, dowdy2018bounds,kuntz2019bounding, sakurai2022interval, sakurai2018bounding, dowdy2018dynamic, holtorf2024tighter}, but they become computationally demanding for transient analysis due to the combinatorial growth in the number of moment variables.
Furthermore, these existing approaches have inherent limitations in that they require solving the underlying equations or optimization problems repeatedly for different initial conditions (probability distributions). 
Thus, a key challenge is to develop methods that enable efficient and reliable analysis of transient statistics while avoiding recomputation across different initial conditions.

In this study, we address this challenge by leveraging the dual representation of the CME, namely the Kolmogorov's backward equation \cite{norris1998markov}, to develop a method for bounding transient statistics of molecular copy numbers.
The proposed formulation avoids the non-closure issue of moment equations by shifting the source of infinite-dimensionality from the moment hierarchy to the dependence on the initial state. 
This enables (i) the construction of finite-dimensional state-space models that provide upper and lower bounds on transient moments by exploiting the monotonicity of the CTMC generator, and (ii) computing moments for different initial conditions by simple inner-product operations with the initial distribution. 
Therefore, transient moment bounds can be computed for different initial distributions with significantly reduced computational cost.
In particular, for certain classes of SRNs, the bounding ODEs can be constructed in a fully constructive manner, providing a systematic approach for explicit computation of rigorous bounds.
The proposed method can be regarded as a dual counterpart of the FSP \cite{munsky2006finite} in that both provide guaranteed approximations using a state-space truncation approach, but the former provides bounds on moments whereas the latter computes probability distributions of the CME. 
The proposed method is aimed at directly computing rigorous bounds on transient moments, which are not directly provided by FSP. Moreover, once the bounding system is constructed, the bounds can be efficiently evaluated for multiple initial distributions without recomputation.

The rest of this paper is organized as follows. Section \ref{sec:modeling} introduces the model of SRNs and their moment dynamics. Section \ref{sec:main result} presents a dual representation of the CME, namely the Kolmogorov's backward equation, and develops the bounding ODEs. Section \ref{sec:exaple} demonstrates the proposed approach through two numerical examples with different classes of propensity functions. Finally, Section \ref{sec:conclusion} concludes the paper.

\noindent
\textbf{Notations:}
The symbol $\mathbb{P}(\cdot)$ is a probability mass function, and 
$\mathbb{E}[f(\cdot)]$ is the expectation of the mass function defined by $\sum_{\bm{x}} f(\bm{x}) \mathbb{P}(\bm{x})$.
The symbol $\mathrm{supp}(\cdot)$ denotes the support of a probability distribution. 
The expression $\bm{x}^{\bm{\alpha}}$ with $\bm{x}=[x_1, x_2, \ldots, x_n]^\top$ and $\bm{\alpha}=[\alpha_1, \alpha_2, \ldots, \alpha_n]^\top \in \mathbb N_0^n$ denotes a multivariate monomial of $x_1^{\alpha_1} x_2^{\alpha_2} \dots x_n^{\alpha_n}$, and $|\bm \alpha| := \sum_{i} \alpha_i$ is the order of monomial or moment.
For vectors $\bm{x}$ and $\bm{y}$, $\bm x \leq \bm y$ denotes the component-wise inequality.
The symbols $\qty{\cdot}^{\pm}$ denote the positive and negative parts, i.e., $\qty{a}^{+}\coloneq \max\qty(a,0),\ 
\qty{a}^{-} \coloneq \min\qty(a,0)$, respectively.


\section{Modeling of Stochastic Reaction Networks and Problem Statement}\label{sec:modeling}

We consider a stochastic reaction network (SRN) composed of $n$ molecular species $M_i \:(i = 1, 2, \dots, n)$ and $r$ reaction species $R_j \:(j = 1, 2, \dots, r)$. 
Let $x_i$ denote the copy number of species $M_i$, and define the state of the SRN as $\bm x\coloneq [x_1,x_2,\dots,x_n]^\top \in \mathcal{X} \subseteq \mathbb N_0^n$. 
Here, we are interested in the stochastic dynamics of the state $\bm{x}$ induced by reaction kinetics.

The reaction species $R_j$ is characterized by a stoichiometry vector $\bm s_j \coloneq \qty[s_{j,1},s_{j,2},\dots,s_{j,n}]^\top \in \mathbb Z^n$ describing the change in $\bm x$ upon the occurrence of the reaction, \textit{i.e.,} $\bm x \to \bm x+\bm s_j$, and a propensity function $\lambda_j(\bm x)$ representing the probability of its occurrence per unit time. 
The functional form of $\lambda_j(\bm x)$ may vary depending on the context and type of reaction kinetics (e.g., mass-action, Hill, Michaelis–Menten, etc.). 
Consequently, since the state transition $\bm x \to \bm x+\bm s_j$ occurs with the state-dependent rate $\lambda_j(\bm x)$, the SRN can be modeled as a continuous-time Markov chain (CTMC) \cite{anderson2011continuous}. 
Let $\qty{\bm X(t)}_{t \geq 0}$ denote the stochastic process representing the state of the SRN at time $t$. 
Then, the time evolution of the state probability, or the copy number distribution, $p(\bm x,t;p_0) \coloneq \mathbb P(\bm X(t)=\bm x | \bm X(0)\sim p_0)$ is governed by the following Chemical Master Equation (CME) \cite{gillespie1992rigorous}
\begin{align}\label{eq:CME}
    \!\!
    \frac{d}{dt}p(\bm x,t) \!=\!  \sum_{j=1}^{r} \qty{\lambda_j(\bm x\!-\!\bm s_j) p(\bm x\!-\!\bm s_j,t)
    \!-\! \lambda_j(\bm x)p(\bm x,t)}, \!\!
\end{align}
defined for each $\bm{x} \in \mathcal{X}$, where $p_0(\bm x)$ is the probability distribution of the initial state $\bm X(0)$. 
For notational simplicity, we omit the dependence on $p_0$ and write $p(\bm x,t)$ instead of $p(\bm x,t;p_0)$ in the CME \eqref{eq:CME}.
The CME \eqref{eq:CME} is a linear ordinary differential equation (ODE), but, in many practical cases, it becomes infinite-dimension, \textit{i.e.,} $\mathcal{X} = \mathbb{N}_0^n$, unless the state space is bounded by conservation relations of molecular copy numbers.

In what follows, we consider the moments of $p(\bm{x}, t)$, which are intuitive indicators for understanding stochastic behavior of SRNs. 
In general, the time evolution of $\mathbb E\qty[f(\bm X(t))]$ is governed by the following equation derived from the CME \eqref{eq:CME}
\begin{align}\label{eq:ME}
    \!\!
    \frac{d}{dt} \mathbb E\qty[f(\bm X(t))] \!=\!  \sum_{j=1}^{r} \mathbb E\qty[\lambda_j(\bm X(t))\qty{f(\bm X(t)\!+\!\bm s_j)
    \!-\! f(\bm X(t))}], \!\!
\end{align}
where $f(\bm{X})$ is a given function \cite{anderson2011continuous}. 
When $f(\bm{X})=\bm{X}^{\bm \alpha}$, eq. \eqref{eq:ME} is called moment equation.
However, the moment equation \eqref{eq:ME} forms an infinite-dimensional ODE when the SRN contains bimolecular and higher-order reactions, or reactions with Hill- or Michaelis--Menten-type propensity functions, leading to an unclosed hierarchy of moment equations. 
This makes the equation analytically intractable, and obtaining an exact solution often becomes difficult.

Hence, this paper considers the bounding problem for the moments. 
More specifically, we propose a computationally efficient approach to computing the transient upper and lower bounds on $\mathbb{E}[\bm{X}^{\bm{\alpha}}(t)]$ for a given initial distribution $p_0$. 
To this end, we first present a key theoretical idea for a general test function $f(\cdot)$ in Section \ref{subsec:duality} and \ref{subsec:bounding}. 
Then, in Section \ref{subsec:input}, we focus on the case of $f(\bm{X}) = \bm{X}^{\bm{\alpha}}$ to derive a procedure for computing moment bounds.

In what follows, we assume that the CTMC is well-defined and that the quantities of interest are finite.

\begin{assumption}\label{assumption1}
(i) The CTMC $\qty{\bm X(t)}_{t\geq 0}$ is non-explosive on $[0,T]$, and 
(ii) $\mathbb E\qty[|f(\bm X(t))|]<\infty$ for all $t \in [0,T]$. 
\end{assumption}


\section{Main Result}\label{sec:main result}
In this section, we first introduce a dual form of the CME, which is the Kolmogorov's backward equation \cite{norris1998markov}. 
This formulation enables the efficient computation of the moment bounds by restricting the dynamics to a finite-dimensional state-space model, thereby circumventing the infinite-dimensional issues of the moment equation.

\subsection{Dual system for expectation computation}\label{subsec:duality}

Let the operator $\mathcal{L}^\dagger$ acting on functions $g: \mathcal{X} \rightarrow \mathbb{R}$ be defined by 
\begin{align}
    (\mathcal L^\dagger g)(\bm x) \coloneq \sum_{j=1}^{r}\left[\lambda_j(\bm x - \bm{s}_j){g(\bm x - \bm s_j)- \lambda_j(\bm{x})g(\bm x)} \right].
\end{align}
The CME \eqref{eq:CME} and the expectation $\mathbb{E}[f(\bm{X}(t))]$ of interest 
can then be written as 
\begin{align}\label{eq:CME op}
&\frac{d}{dt}p(\cdot,t)=\mathcal L^\dagger p(\cdot,t), \qquad
p(\cdot,0)=p_0, \\
&\mathbb E[f(\bm X(t))]
    = \langle f, p(\cdot,t)\rangle
    = \langle f, e^{t\mathcal L^\dagger} p_0 \rangle,
\end{align}
where the pairing is defined by $\langle f,p\rangle := \sum_{\bm x\in\mathcal X} f(\bm x)p(\bm x).$
Define 
an adjoint operator $\mathcal{L}$ of $\mathcal{L}^\dagger$ with respect to this pairing. 
Then, $\mathbb E[f(\bm X(t))]$ can be equivalently expressed using $\mathcal{L}$ as 
\begin{align}\label{eq:dual equation}
\mathbb E[f(\bm X(t))]
 = \langle f, e^{t\mathcal L^\dagger} p_0 \rangle
    = \langle e^{t\mathcal{L}} f, p_0 \rangle.
\end{align}
The operator $\mathcal{L}$ can be specifically written as 
\begin{align}\label{eq:L}
(\mathcal L g)(\bm x)
\coloneq
\sum_{j=1}^{r}\lambda_j(\bm x)\{g(\bm x+\bm s_j)-g(\bm x)\}. 
\end{align}
Moreover, under Assumption \ref{assumption1}, eq. \eqref{eq:dual equation} implies that the conditional expectation 
$q(\bm x,t)\coloneq \mathbb E[f(\bm X(t))\mid \bm X(0)=\bm x]$ 
satisfies
\begin{align}\label{eq:KBE op}
\frac{d}{dt}q(\cdot,t)=\mathcal L q(\cdot,t),
\qquad
q(\cdot,0)=f, 
\end{align}
which is known as the Kolmogorov's backward equation\cite{norris1998markov}.

Thus, $\mathbb{E}[f(\bm{X}(t))]$ can be represented by dual linear systems given by the state space equations  \eqref{eq:CME op} and \eqref{eq:KBE op}, and the output equation \eqref{eq:dual equation}. 
An advantage of using the dual form \eqref{eq:KBE op} is that, unlike the moment equation \eqref{eq:ME}, it is closed with respect to the order of the function $f$, when $f(\bm{X}) = \bm{X}^{\bm{\alpha}}$. Thus, the equation does not suffer from the unclosed hierarchy issues.
Moreover, the moment $\mathbb E\qty[f(\bm X(t))]$ can be expressed as a linear functional of the initial distribution $p_0$ as shown by the last term in equation \eqref{eq:dual equation}.
Thus, once $e^{t\mathcal{L}}$ has been computed, $\mathbb E\qty[f(\bm X(t))]$ for different initial distributions $p_0$ can be evaluated efficiently since the computation reduces to a simple inner product of $e^{t\mathcal{L}}$ and $p_0$. 
This is in contrast to the CME and the moment equation based approach, where the ODE solutions need to be recomputed for each initial distribution $p_0$ or $\mathbb{E}[f(\bm{X}(0))]$.

Motivated by this observation, we introduce the following assumption on the initial distribution $p_0$.

\begin{assumption}\label{assumption2}
    The support of the initial distribution, denoted by $\mathrm{supp}\, p_0(\bm x)$, is finite.
\end{assumption}
Since SRNs in biological systems operate in the regime where the copy number $\bm{x}$ is low, in practice, it is not restrictive to model $\mathrm{supp}\, p_0(\bm x)$ as finite.
The following theoretical development is based on Assumption \ref{assumption2}, but extensions to general cases would be possible in a similar spirit.


\subsection{Bounds on expectation using dual system}\label{subsec:bounding}

\begin{figure}
    \centering
    \includegraphics[width=\linewidth]{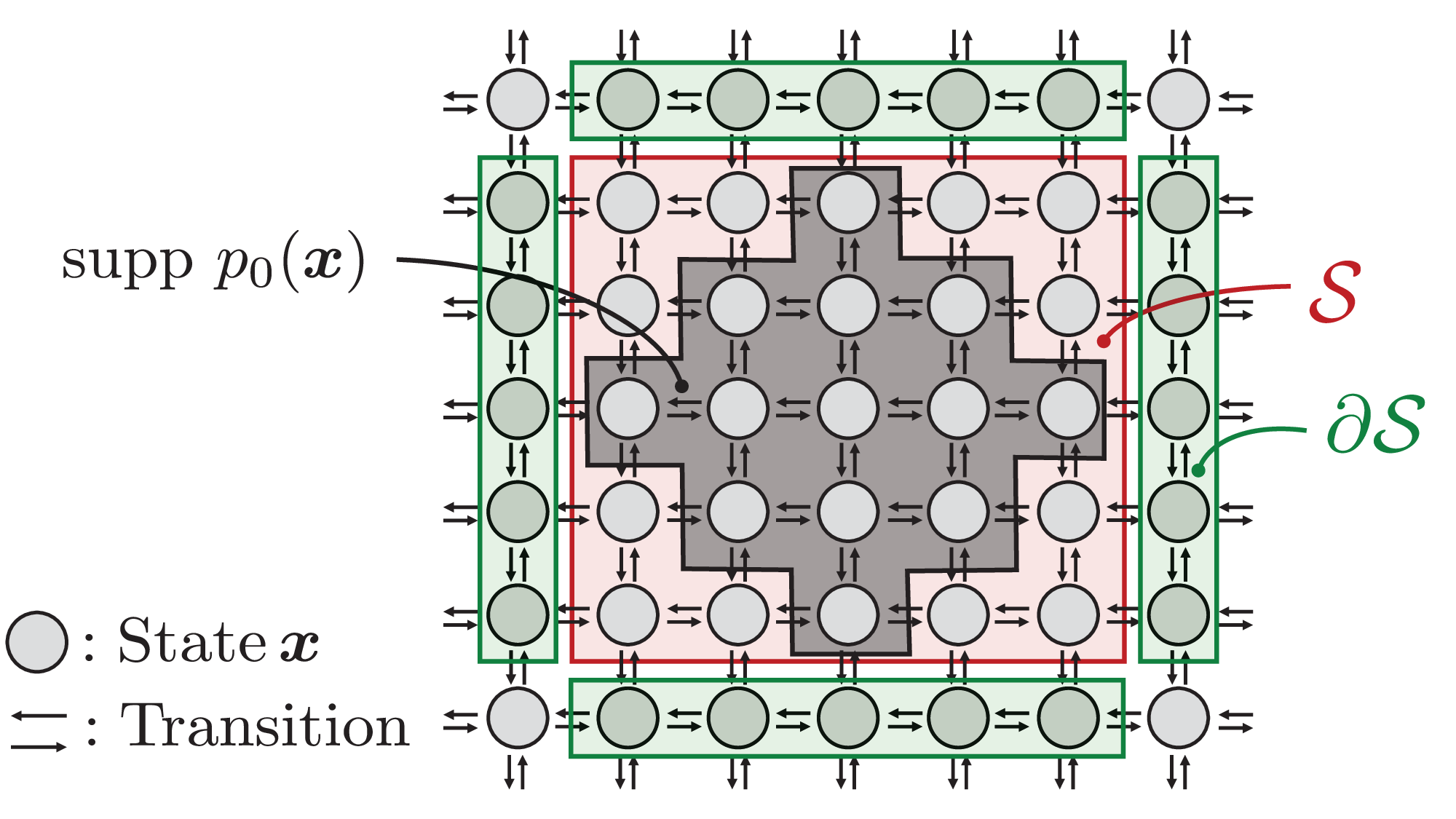}
    \caption{Truncated state space $\mathcal{S}$ and its boundary $\partial \mathcal{S}$. The vector $\bm{q}(t)$ of conditional expectations $\mathbb{E}[f(\bm{X}(t))|\bm{X}(0)=\bm{x}]$ is defined on $\mathcal{S}$ and $\bm{u}(t)$ is  defined on $\partial \mathcal{S}$. 
    Truncating the state space to $\mathcal S$ enables the computation of upper and lower bounds on the expectation using a finite-dimensional system defined on $\mathcal S$.}
    \label{fig:Diagram}
\end{figure}
Based on the observations in the previous subsection, we next derive systems that bound the solution of $\mathbb{E}[f(\bm{X}(t))]$.
Suppose Assumption \ref{assumption2} holds, and consider a truncated state space $\mathcal S \subseteq \mathcal{X}$ with size $|\mathcal S|=N$ such that $\mathrm{supp}\, p_0(\bm x)\subseteq \mathcal S$. Let $\partial \mathcal S$ denote the set of states with size $|\partial \mathcal S|=b$ that are reachable from some state in $\mathcal S$ by a single reaction and are in the complement of $\mathcal S$.
A diagram illustrating these sets is shown in Fig. \ref{fig:Diagram}.
Let $\bm q(t)\in \mathbb R^N$ be the vector that stacks the conditional expectation $q(\bm x,t)$ over $\bm x\in\mathcal S$, and let $\bm p_0\in \mathbb R_+^N$ be the vector that stacks the initial distribution $p_0(\bm x)$ over $\bm x\in\mathcal S$.

Then, eq. \eqref{eq:KBE op} implies that the time evolution of $\mathbb E\qty[f(\bm X(t))]$ can be represented by the following $N$-dimensional state-space model
\begin{align}
\mathcal B\ \colon \
&\begin{cases}
\displaystyle \frac{d}{dt}\bm q(t) = L_{\mathcal S}\, \bm q(t) + L_{\partial\mathcal S}\, \bm u(t)\\
y(t) = \bm p_{0}^\top \bm q(t),
\end{cases}
\end{align}
where $L_{\mathcal S}\in \mathbb R^{N\times N}$ is the matrix corresponding to the operator $\mathcal L$ representing the transitions within the truncated state space $\mathcal S$, and $L_{\partial \mathcal S}\in \mathbb R_+^{N\times b}$ 
is a matrix corresponding to the transitions from $\mathcal S$ to the boundary set $\partial \mathcal S$.
The output $y$ represents $\mathbb E\qty[f(\bm X(t))]$, which follows from 
eq. \eqref{eq:dual equation}, and the input $\bm u(t)\in \mathbb R^{b}$ is the vector that stacks the conditional expectation $q(\bm x,t)$ over $\bm x\in \partial \mathcal S$.

However, a key difficulty in evaluating $y(t)$ in the system $\mathcal{B}$ is that $\bm u(t)$ depends on the conditional expectations associated with the boundary states $\partial \mathcal S$. Therefore, it is necessary to characterize the possible values of $\bm u(t)$. The following lemma provides a key observation for our theoretical development.

\begin{lemma}\label{lemma}
  Let $(\bm{u}_1(t), \bm{q}_1(t), y_1(t))$ and $(\bm{u}_2(t), \bm{q}_2(t), y_2(t))$ denote the input, state, and output of the state-space model $\mathcal B$, respectively. 
  If $\bm q_1(0)=\bm q_2(0)$, and $\bm u_1(t)\leq \bm u_2(t)$ for all $t\in[0,T]$, then
  \begin{align}
    \bm q_1(T) \leq  \bm q_2(T),\quad  y_1(T) \leq  y_2(T),
  \end{align}
for any given $T$.
\end{lemma}

This lemma shows that the system $\mathcal B$ is monotone with respect to the input $\bm{u}(t)$.
Therefore, by introducing $\bm u^{+}(t)\in \mathbb R^b$ and $ \bm u^{-}(t)\in \mathbb R^b$ satisfying 
\begin{align}\label{eq:inequality1 of input}
  \bm u^{-}(t) \leq \bm u(t) \leq \bm u^{+}(t),\ \forall t\in[0,T],
\end{align}
we can define two state-space models $\mathcal B_+$ and $\mathcal B_-$ giving the bounds of the expectation $\mathbb{E}[f(\bm{X}(t))]$ as  
\begin{align} 
\mathcal B_{+}\ \colon \
&\begin{cases}
\displaystyle\frac{d}{dt}\bm q^+(t) = L_{\mathcal S}\, \bm q^+(t) + L_{\partial\mathcal S}\, \bm u^{+}(t)\\
y^+(t) = \bm p_{0}^\top \bm q^+(t), \label{eq:B+}
\end{cases}
\\[0.8em]
\mathcal B_{-} \ \colon \
&\begin{cases}
\displaystyle \frac{d}{dt}\bm q^-(t) = L_{\mathcal S}\, \bm q^-(t) + L_{\partial\mathcal S}\, \bm u^{-}(t)\\
y^-(t) = \bm p_{0}^\top \bm q^-(t),\label{eq:B-}
\end{cases}
\end{align}
with $\bm q^+(0)=\bm q^-(0)=\bm q(0)$. 
This is more formally summarized in the following theorem.

\begin{theorem}\label{theorem1}
Consider an SRN with stoichiometric vectors 
$\{\bm{s}_i\}_{i=1}^{r}$ and propensity functions $\{\lambda_i(\bm{x})\}_{i=1}^{r}$.
Let $\bm{X}(t)$ denote the corresponding stochastic process, and let 
$\mathbb E[f(\bm X(t))]$ be the expectation of a test function $f(\cdot)$.
Suppose Assumption \ref{assumption1} and \ref{assumption2} hold, and define the state-space models $\mathcal{B}_+$ and $\mathcal{B}_-$ satisfying \eqref{eq:inequality1 of input}, respectively. 
Then, for all $t \in [0, T]$, the expectation $\mathbb E\qty[f(\bm X(t))]$ satisfies
\begin{align}
    y^-(t) \leq \mathbb E\qty[f(\bm X(t))] \leq y^+(t).
\end{align}
\end{theorem}

Theorem \ref{theorem1} shows that the upper and lower bounds on the transient expectation $\mathbb E\qty[f(\bm X(t))]$ can be obtained by computing the outputs of the two state-space models $\mathcal B_+$ and $\mathcal B_-$, respectively.
Hence, the moment bounding problem reduces to the construction of the input bounds, $\bm u^{+}(t)$ and $\bm u^{-}(t)$, that satisfy eq. \eqref{eq:inequality1 of input}.


\subsection{Bounds on moments via input bounding}\label{subsec:input}

In this section, we restrict the class of the test functions to monomials, \textit{i.e.,} $f(\bm{X}) = \bm{X}^{\bm{\alpha}}$, to focus on the moment bounding problem. 
We then show a method for obtaining $\bm u^{+}(t)$ and $\bm u^{-}(t)$ that satisfy inequality \eqref{eq:inequality1 of input}. 

Since $\bm u(t)$ is the vector obtained by stacking the conditional moments $q(\bm x,t)$ for all $\bm x \in \partial \mathcal S$, it suffices to consider the bounds of these conditional moments.
When $f(\bm{X}) = \bm{X}^{\bm{\mu}}$, let the $i$-th components of $\bm u(t)$, $\bm u^{+}(t)$ and $\bm u^{-}(t)$ be denoted by $u_{\bm \mu,i}(t)$, $u_{\bm \mu,i}^{+}(t)$ and $u_{\bm \mu,i}^{-}(t)\in\mathbb R$, respectively.
The following lemma provides a way to compute the upper and lower bounds on the dynamic conditional moments $\mathbb{E}[\bm{X}^{\bm{\alpha}}(t) | \bm{X}(0) = \bm{x}]$, \textit{i.e.,} $u_{\bm \alpha,i}^{+}(t)$ and $u_{\bm \alpha,i}^{-}(t)$, respectively.

\begin{lemma}\label{lemma2}
Consider the bounding systems $\mathcal{B}_+$ and $\mathcal{B}_{-}$ defined in eq. \eqref{eq:B+} and eq. \eqref{eq:B-}.
Suppose $\mathbb E\qty[|\bm X^{\bm \mu}(t)|\mid \bm X(0)=\bm x]<\infty$ for all $\bm{\mu}\in \mathbb N_0^n$ and for all $\bm x\in \mathcal X$.
If, for each $\bm{\mu}$ satisfying $|\bm{\mu}| \le |\bm\alpha|$, there exist $|\bm{\alpha}|$-th degree polynomials
\begin{align}
  h_{\bm \mu}^+(\bm x):=\!\sum_{|\bm\nu| \leq |\bm\alpha|} c_{\bm\mu,\bm\nu}^+\,\bm x^{\bm\nu},\ \ 
  h_{\bm \mu}^-(\bm x):=\!\sum_{|\bm\nu| \leq |\bm\alpha|} c_{\bm\mu,\bm\nu}^-\,\bm x^{\bm\nu},
\end{align}
such that 
\begin{align}\label{eq:inequality of operator}
  h_{\bm\mu}^-(\bm x)\ \le\ \bigl(\mathcal L\,\bm x^{\bm\mu}\bigr)(\bm x)\ \le\ h_{\bm\mu}^+(\bm x),
\end{align}
where $c_{\bm\mu,\bm\nu}^+\in \mathbb R$ and $c_{\bm\mu,\bm\nu}^- \in \mathbb R$ are constants.
Then, for $|\bm{\mu}| \le |\bm\alpha|$, the time evolutions of $u_{\bm\mu,i}^{+}(t)$ and $u_{\bm\mu,i}^{-}(t)$ are governed by the following ODEs

\begin{align}
  \frac{d}{dt}u_{\bm\mu,i}^{\pm}(t)
  =&\ c_{\bm\mu,\bm\mu}^{\pm}\,u_{\bm\mu,i}^{\pm}(t) \notag\\
  &+\sum_{\substack{|\bm\nu| \leq |\bm\alpha| \\ \bm\nu \neq \bm\mu}}\!
  \Bigl(\qty{c_{\bm\mu,\bm\nu}^{\pm}}^+\,u_{\bm\nu,i}^{\pm}(t)
        +\qty{c_{\bm\mu,\bm\nu}^{\pm}}^-\,u_{\bm\nu,i}^{\mp}(t)\Bigr),
  \label{eq:ODE of u^pm}
\end{align}
where $u_{\bm\mu,i}^{+}(0)=u_{\bm\mu,i}^{-}(0)=\bm x_i^{\bm\mu}$ with $\bm x_i$ being the state on $\partial \mathcal S$ corresponding to the $i$-th component of $\bm u(t)$, and
$u_{\bm 0,i}^{+}(t)=u_{\bm 0,i}^{-}(t)=1$.
The paired signs $\pm$ and $\mp$ are taken in the same order.
\end{lemma}

This lemma implies that $u_{\bm\alpha,i}^{+}(t)$ and $u_{\bm\alpha,i}^{-}(t)$ can be obtained by the coupled linear ODEs.
In particular, the problem of computing upper and lower bounds on the conditional moments $\mathbb E[\bm{X}^{\bm{\alpha}}(t)|\bm{X}(0) = \bm{x}]$ reduces to the problem of finding the polynomials $h_{\bm \mu}^+(\bm x)$ and $h_{\bm \mu}^-(\bm x)$ that satisfy inequality \eqref{eq:inequality of operator}.
Note that even if a polynomial $h_{\bm \mu}^-(\bm x)$ providing a lower bound cannot be obtained, we may simply set $\bm u_{\bm \mu,i}^-(t)=0$ due to the nonnegativity of the moments.
Therefore, once a polynomial $h_{\bm \mu}^+(\bm x)$ is found, upper and lower bounds on the moment $\mathbb E[\bm{X}^{\bm{\alpha}}(t)]$ can be derived from Theorem \ref{theorem1}.

When all reactions in an SRN belong to a certain class of elementary reactions \cite{khammash2022cybergenetics}\footnote{Elementary reactions refer to the zero-th, first, and second order reactions, where the propensity functions $\lambda_i(\bm{x})$ are given either by a constant $k$, a linear function $kx_i$, or quadratic functions $k x_i x_j$ or $kx_i(x_i-1)$, where $k$ is a constant, and $x_i$ and $x_j$ are the copy numbers of reactant molecular species.}, 
the bounding function $h_{\bm{\mu}}^+(\bm{x})$ can be analytically obtained using the stoichiometric vector $\{\bm{s}_i\}_{i=1}^{r}$ and the propensity functions $\{\lambda_i(\bm{x})\}_{i=1}^{r}$.

\begin{theorem}\label{theorem2}
Consider an SRN with stoichiometric vectors $\{\bm{s}_i\}_{i=1}^{r}$ and propensity functions $\{\lambda_i(\bm{x})\}_{i=1}^{r}$. Suppose reactions in the SRN satisfy the following properties: (i) all reactions are elementary, and (ii) $\bm s_j \leq \bm 0$ for all $j \in \mathcal{J}_{\mathrm{bi}}$, where $\mathcal{J}_{\mathrm{bi}} \subseteq \{1,2,\ldots,r\}$ is the set of indices of bimolecular reactions. 
The following $h_{\bm \mu}^+(\bm x)$ satisfies inequality \eqref{eq:inequality of operator}
\begin{align}\label{eq:h+}
    h_{\bm \mu}^+(\bm x) =& \sum_{j \notin \mathcal J_{\mathrm{bi}}}\sum_{i=1}^{n} \mu_i s_{j,i}\, \lambda_j(\bm x)\bm x^{\bm \mu-\bm e_i} \notag \\
    &+ \sum_{j=1}^{r}\sum_{\substack{\bm \nu \le \bm \mu\\ |\bm \nu|\ge 2}}\binom{\bm \mu}{\bm \nu}\bm s_j^{\bm \nu}\, \lambda_j(\bm x)\bm x^{\bm \mu-\bm \nu}, 
\end{align}
where $\bm{e}_i$ is a $n$-dimensional vector whose $i$-th component is 1 and the other components are zero, and $\binom{\bm{\mu}}{\bm{\nu}}\coloneq\prod_{i=1}^n \binom{\mu_i}{\nu_i}$ for 
 multi-indices $\bm\mu$ and $\bm\nu$ with $\bm\nu\le \bm\mu$. 
\end{theorem}

These analytic expressions allow for the systematic construction of the upper bound function $h_{\bm{\mu}}^{+}(\bm{x})$ in Lemma \ref{lemma2}, which 
leads to the computation of the moment bounds $\mathbb{E}[\bm{X}^{\bm{\alpha}}(t)]$ in Theorem \ref{theorem1}. 
The assumption (ii) in Theorem \ref{theorem2} means that any bimolecular reaction does not increase the copy number of any molecular species in the SRN.
Specifically, it corresponds to reactions in which two molecules, $M_i$ and $M_j$, bind and are then degraded or inactivated, such as $M_i+M_j\rightarrow \emptyset$. 
Due to this restriction, Theorem 2 does not provide a constructive characterization of $h_{\bm{\mu}}^{+}(\bm{x})$ for all classes of elementary reactions. Nevertheless, it remains applicable to many practically relevant reaction classes, including the antithetic integral feedback (AIF) motif used as a molecular feedback controller \cite{briat2016antithetic}.

\begin{remark}\label{remark2}
The analytic expression in Theorem 2 is also useful for SRNs that include rational propensity functions if these functions can be bounded by one of the functional forms in elementary reactions. For example, the upper bound of Hill functions can be written by a constant as ${K^\eta}/{(K^\eta + x^{\eta})} \le 1$, where $K$ is a Hill constant and $\eta$ is a Hill coefficient. The constant in the right-hand side corresponds to the function form of zero-th order reactions. 
We can then use this constant as a proxy function in eq. \eqref{eq:h+} to obtain $h^+_{\bm{\mu}}(\bm{x})$ in 
Theorem \ref{theorem2}.
\end{remark}


\section{Numerical Examples}\label{sec:exaple}
In this section, we demonstrate the proposed method and illustrate its effectiveness.
Specifically, in Section \ref{subsec:elementary reactions example}, we apply the proposed method to an SRN with elementary reactions and show that the gap between the upper and lower bounds decreases as the truncated state space $\mathcal S$ expands.
In Section \ref{subsec:Hill example}, we then apply the proposed method to an SRN with Hill-type propensity functions, thereby illustrating its applicability to SRNs with rational propensity functions.

\subsection{SRN with elementary reactions}\label{subsec:elementary reactions example}

\begin{table}[tb]
  \caption{Propensity functions, stoichiometric vectors, and rate constants for the dimerization reaction network}
  \label{tab:list of p&s of dimer}
  \centering
  \begin{tabular}{cccc}
    \hline
    Reaction $R_j$  & $\lambda_j(\bm x)$  &  $\bm s_j$ & $\theta_j$ \\
    \hline \hline
     $R_1:\emptyset \rightarrow M \ $ & $\theta_1$ & $+1$ & $5.0$\\
     $R_2:M \rightarrow \emptyset \ $ & $\theta_2x$  & $-1$ & $\mathrm{ln}(2)/20$\\
     $R_3:2M \rightarrow \emptyset$ & $\theta_3x(x-1)$  & $-2$ & $0.02$\\
    \hline
  \end{tabular}
\end{table}

Consider the dimerization reaction network consisting of $r=3$ reactions with propensity functions $\lambda_j(\bm x)$, stoichiometric vectors $\bm s_j$ and reaction rate constants $\theta_j$ listed in Table \ref{tab:list of p&s of dimer}.
The state of the SRN is defined by the copy number $x$ of the molecule $M$. 
Our goal is to compute the guaranteed bounds of the mean $\mathbb E\qty[X(t)]$ and the variance $\mathbb V\qty[X(t)]\coloneq \mathbb E\qty[X^2(t)]-\mathbb (E\qty[X(t)])^2$ of the copy number $X(t)$.

We define the truncated state space by $\mathcal S = \{x \in \mathbb N_0 $ $\mid 0 \leq x \leq N-1\}$ with size $|\mathcal S|=N$. 
Let $\bm q(t)$ be the vector that stacks conditional moment $\mathbb E\qty[X^\alpha(t)\mid X(0)= x],\alpha=1,2$. 
Then, the input term of the state-space model $\mathcal B$ is given by
\begin{align}
    L_{\partial\mathcal S}\,\bm u(t) = \qty[0,0,\dots,0,\theta_1u_{\alpha,1}(t)]^\top \in \mathbb R_+^{N},
\end{align}
Therefore, it suffices to obtain upper and lower bounds on $u_{\alpha,1}(t)$. 
Here, we specifically describe how to obtain the bounds for $\alpha=1$, while those for $\alpha=2$ can be obtained in the same manner.
First, we set $u_{1,1}^-(t)=0$ due to the nonnegativity of the moments, and derive the upper bound $u_{1,1}^+(t)$ by the method presented in Section \ref{subsec:input}.
By Theorem \ref{theorem2}, the polynomial $h_1^+(x)$ is given by $h_1^+(x) = \theta_1 - \theta_2x$.
Therefore, by Lemma \ref{lemma2}, the upper bound $u_{1,1}^+(t)$ for $\alpha=1$ satisfies the following ODE
\begin{align}
    \frac{d}{dt}u_{1,1}^+(t) =& - \theta_2\,u_{1,1}^+(t) + \theta_1, \label{eq:u1+ODE of dimer} 
\end{align}
where the initial condition is given by $u_{1,1}^+(0)=N$.
Solving the ODE \eqref{eq:u1+ODE of dimer}, we obtain $u_{1,1}^+(t)$, which is then substituted into the bounding systems $\mathcal B_+$ and $\mathcal B_-$.
The outputs $y^{+}(t)$ and $y^{-}(t)$ of these bounding systems give 
the upper and lower bounds on $\mathbb E\qty[X(t)]$ as shown in Theorem \ref{theorem1}.
The bounds on the second moment $\mathbb E\qty[X^2(t)]$ are obtained in the same manner.

\begin{figure}
    \centering
    \includegraphics[width=\linewidth]{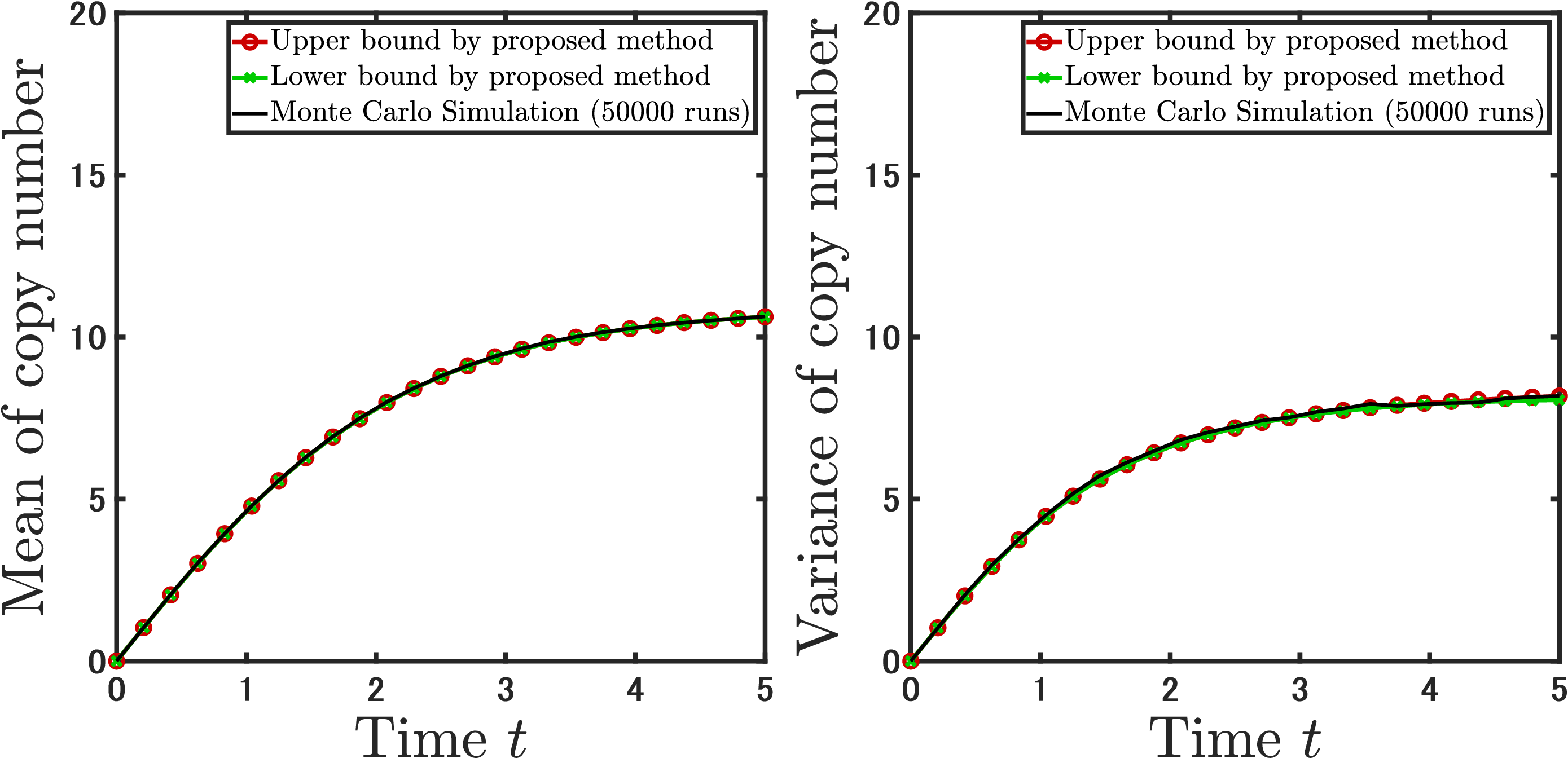}
    \caption{Upper and lower bounds on mean $\mathbb{E}[X]$ (left) and variance $\mathbb{V}[X]$ (right) of the molecular copy number for the dimerization reaction network, compared with Monte Carlo simulations. The truncated state-space size is $N=25$, the initial condition is $X(0)=0$, and the Monte Carlo results are obtained from 50000 trajectories. The Monte Carlo results lie within the computed bounds, confirming the validity of the proposed bounding approach.}
    \label{fig:results of 1D}
\end{figure}

\begin{figure}
    \centering
    \includegraphics[width=0.9\linewidth]{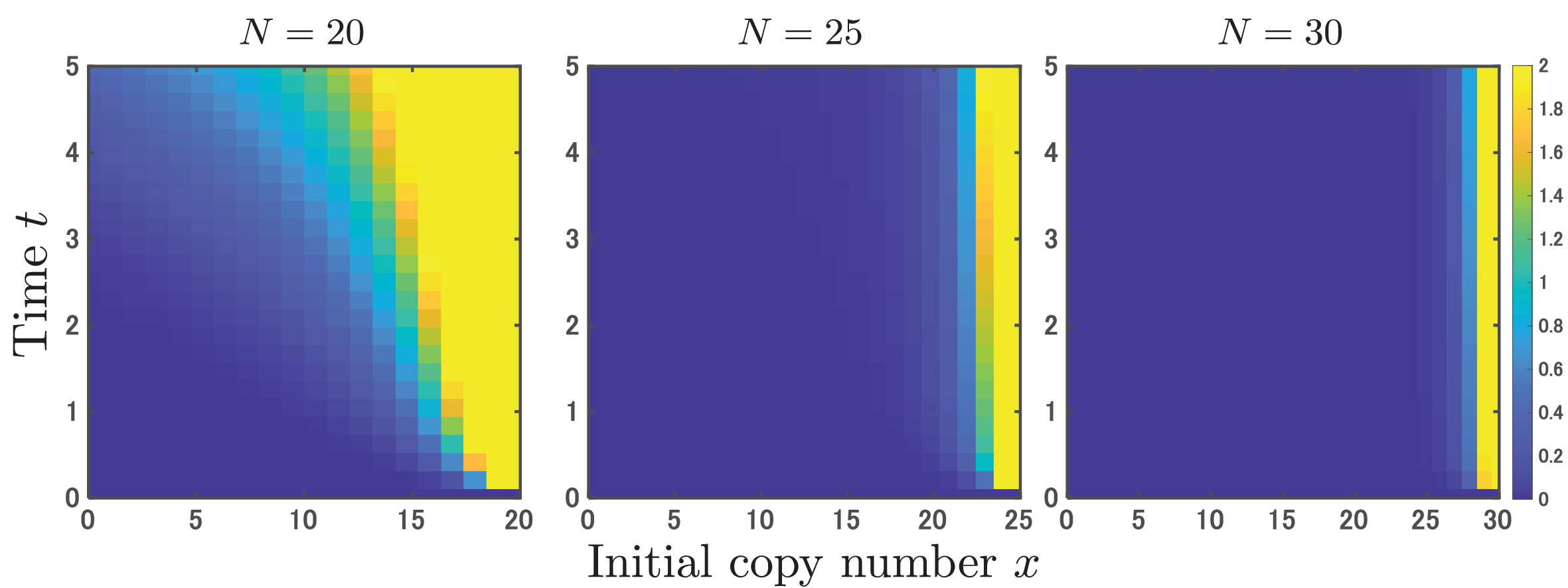}
    \caption{Gap between the upper and lower bounds of the mean $\mathbb{E}[X]$ for state space sizes $N=20,\ 25,$ and $30$. The gap decreases as $N$ increases, showing that expanding the truncated state space yields tighter bounds.}
    \label{fig:error of dimer}
\end{figure}

Figure \ref{fig:results of 1D} shows the upper and lower bounds on 
mean $\mathbb E\qty[X(t)]$ and variance $\mathbb V\qty[X(t)]$ for $N=25$ obtained by the proposed approach, and those obtained by 50000 Monte Carlo simulations based on the stochastic simulation algorithm \cite{gillespie1976general}. 
In this example, the initial copy number is set to zero, which corresponds to $p_0(0)=1$ and $\bm{p}_0 = [1,0,\ldots,0]^\top$ in $\mathcal{B}_+$ and $\mathcal{B}_{-}$.
Figure \ref{fig:results of 1D} shows that the proposed approach yields valid bounds for all $t \ge 0$. 
The bounds tend to become looser over time. Nevertheless, the gap becomes smaller as it recovers truncated states. 
Thus, even over long time horizons, the deterioration in bound tightness can be mitigated by appropriately expanding $\mathcal S$.
To see this, we compute the gap $\bm e(t)\coloneq \bm q^+(t)-\bm q^-(t)$ for truncated state space $\mathcal S$ with various sizes $N (=|\mathcal{S}|)$. 
The gap for the mean $\mathbb E\qty[X(t)]$ is shown in Fig. \ref{fig:error of dimer}, where the size of the state space is set $N=20,\ 25,$ and $30$.

A key advantage of the proposed method is that it can obtain moment bounds for various initial distributions or initial states in a single computation, thereby substantially reducing the computational cost.
For comparison with the optimization-based approach, we use the optimization-based approach \cite{sakurai2018bounding} as a benchmark.
For the $25$ time points up to $T=5.0$ shown in Fig. \ref{fig:results of 1D}, the computation times were $1316$ s for the benchmark method and $0.2$ s for the proposed method. 
Notably, the reported computation time for the proposed method includes the simultaneous computation of solutions for all initial states $X(0)=0,1,\dots,24$, rather than only those displayed in Fig. \ref{fig:results of 1D}. Therefore, the proposed method can substantially reduce the computational cost when moment bounds for multiple initial states are required.
The computation was performed by MATLAB R2025b on Intel Xeon W-2265 processor with 256 GB RAM.

\subsection{SRN with rational propensity functions}\label{subsec:Hill example}

\begin{table}[tb]
  \caption{Propensity functions, stoichiometric vectors, and rate constants in genetic toggle switch}
  \label{tab:list of p&s of toggle}
  \centering
  \begin{tabular}{cccc}
    \hline
    Reaction $R_j$  & $\lambda_j(\bm x)$  &  $\bm s_j$ & $\theta_j$ \\
    \hline \hline
     $R_1:\emptyset \rightarrow A$ & $\frac{K_1}{1+\qty(x_B/\theta_1)^3}$ & $[1,0]^\top$ & $10.0$\\
     $R_2:A\rightarrow \emptyset $ & $\theta_2x_A$  & $[-1,0]^\top$ & $1.0$\\
     $R_3:\emptyset \rightarrow B$ & $\frac{K_2}{1+\qty(x_A/\theta_3)^3}$  & $[0,1]^\top$ & $10.0$\\
     $R_4:B\rightarrow \emptyset $ & $\theta_4x_B$  & $[0,-1]^\top$ & $1.0$\\
    \hline
  \end{tabular}
\end{table}

\begin{figure}
    \centering
    \includegraphics[width=\linewidth]{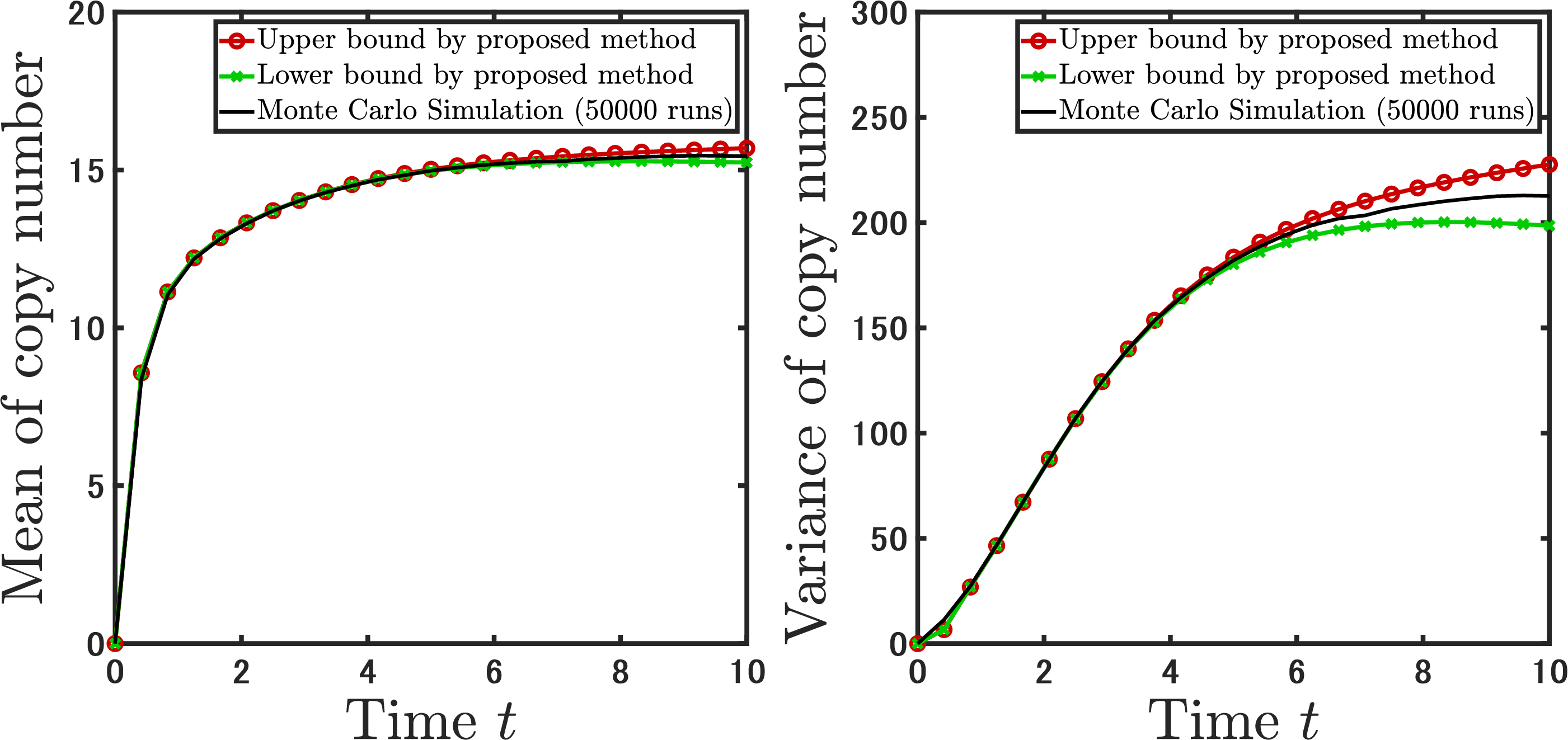}
    \caption{Upper and lower bounds on mean $\mathbb{E}[X_B(t)]$ (left) and variance $\mathbb{V}[X_B(t)]$ (right) of the molecular copy number of protein B in the genetic toggle switch. 
    The truncated state-space size is $(N_A,N_B)=(50,50)$, and the initial condition is $[X_A(0),X_B(0)]^\top=[0,0]^\top$. The results show that the proposed method provides bounds even for rational propensity functions, while the variance bounds are looser than the mean bounds.}
    \label{fig:results of 2D}
\end{figure}

To show an application to SRNs with rational propensity functions, we consider the genetic toggle switch \cite{gardner2000construction}, where two molecular species (proteins) $A$ and $B$ mutually repress each other's production (gene expression). 
This network consists of $r=4$ reactions listed in Table \ref{tab:list of p&s of toggle}, where $R_1$ and $R_3$ correspond to the production regulated by repression, and $R_2$ and $R_4$ represent degradation. 
Let $X_A(t)$ and $X_B(t)$ denote the copy numbers of proteins $A$ and $B$, respectively.
Then, the state of the SRN is defined by $\bm{x}=[x_A,x_B]^\top$, and for each state label $i$, we denote the corresponding state vector by $\bm{x}_i=[x_{i,A},x_{i,B}]^\top$. 
The propensity functions, stoichiometric vectors, and values of the reaction rate constants for each reaction $R_j$ are listed in Table \ref{tab:list of p&s of toggle}.
In addition, the maximum production rates for proteins A and B are set to $(K_1,K_2)=(30,30)$.
For this reaction system, we compute the mean $\mathbb E\qty[X_B(t)]$ and the variance $\mathbb V[X_B(t)]$ of protein B.
To this end, let the truncated state space be $\mathcal S = \qty{\bm x \in \mathbb N_0^2 \mid 0 \leq x_k \leq N_k-1;\ k = A,B}$ with size $|\mathcal S|=N_A\times N_B$.
Here, we specifically describe how to obtain bounds on the mean. 
First, $\qty(\mathcal L\, x_B)(\bm x)$ 
is given by 
\begin{align}
    \qty(\mathcal L\, x_B)(\bm x) = \frac{K_2}{1+\qty(x_A/\theta_3)^3} - \theta_4x_B, 
\end{align}
which includes rational functions due to the Hill function $\lambda_3(\bm{x})$. 
As noted in Remark \ref{remark2}, the Hill function is bounded as $0\leq {K_2}/[1+\qty(x_A/\theta_3)^3] \leq K_2$, and hence
\begin{align}
    - \theta_4x_B \leq \qty(\mathcal L\, x_B)(\bm x) \leq K_2 - \theta_4x_B.
\end{align}
Therefore, Lemma \ref{lemma2} provides the corresponding input bounds
$u^+_{1,i}(t)$ and $u^-_{1,i}(t)$. Substituting them into $B^+$ and $B^-$,
respectively, Theorem 1 gives the upper and lower bounds on
$\mathbb{E}[X_B(t)]$.
Figure \ref{fig:results of 2D} shows the resulting upper and lower bounds on $\mathbb E\qty[X_B(t)]$, together with those on $\mathbb V\qty[X_B(t)]$ obtained in the same manner, for $(N_A,N_B)=(50,50)$ and the initial copy numbers of the proteins given by $[X_A(0),X_B(0)]^\top=[0,0]^\top$.
Figure \ref{fig:results of 2D} shows that upper and lower bounds can be obtained even for SRNs with rational propensity functions.
It can also be observed that the bounds on the variance are looser than those on the mean. This loss of tightness is inherent in the proposed approach, since the upper bound on the variance is constructed by subtracting the lower bound on $\mathbb E[X]^2$ from the upper bound on $\mathbb E[X^2]$, while the lower bound is constructed using the opposite combination.

\section{Conclusion}\label{sec:conclusion}
We have proposed a method for computing upper and lower bounds on transient moments of molecular copy numbers in SRNs with theoretical guarantees based on the Kolmogorov's backward equation.
Specifically, we reduced the problem of bounding transient moments to that of bounding polynomials by exploiting the monotonicity of the CTMC generator.
Consequently, the proposed method enables the computation of transient moment bounds across multiple initial conditions through linear time-invariant (LTI) systems, offering a computationally efficient alternative to existing optimization-based bounding approaches \cite{sakurai2018bounding, dowdy2018dynamic, holtorf2024tighter}.
A fundamental source of conservatism in the proposed method is the construction of the bounding functions in eq. \eqref{eq:inequality of operator}, which are used to obtain the upper and lower bounds on the boundary input.
Developing systematic methods for constructing tighter, ideally optimal, bounding functions and extending the construction to a broader class of reaction networks are important directions for future work.

\textbf{Acknowledgments:} The authors used ChatGPT (\url{https://chatgpt.com}) for language editing,
reviewed and edited its output, and take full responsibility for the manuscript.



\appendix
\subsection{Proof of Lemma \ref{lemma}}\label{app.1}
The matrices $L_{\mathcal S}$ and $L_{\partial \mathcal S}$ in the state-space model $\mathcal B$ have the entries
\begin{equation*}
 \qty[L_{\mathcal S}]_{i,j}=
  \left\{
  \begin{array}{l}
      \begin{aligned} 
          & -\sum\nolimits_{k=1}^{r}\lambda_k(\bm x_i) \quad \mathrm{if}\ \  i = j\\

          &\lambda_k(\bm x_i)  \ \ \quad   \mathrm{for\ all}\ j \ \mathrm{such\ that} \ \bm x_j = \bm x_i+\bm s_k\\

          &0  \ \ \ \quad \qquad \mathrm{otherwise} \\
      \end{aligned}
  \end{array}
  \right.
\end{equation*}
\begin{equation*}
 \qty[L_{\partial \mathcal S}]_{i,j}=
  \left\{
  \begin{array}{l}
      \displaystyle
      \begin{aligned} 
          &\lambda_k(\bm x_i) \ \quad   \mathrm{for\ all}\ j \ \mathrm{such\ that} \ \bm x_j = \bm x_i+\bm s_k\\

          &0  \ \ \quad \qquad \mathrm{otherwise} \\
      \end{aligned}
  \end{array}
  \right.
\end{equation*}
Therefore, since the propensity functions $\lambda_j(\bm x)$ are nonnegative, the matrix $L_{\mathcal S}$ is a Metzler matrix, and the matrix $L_{\partial \mathcal S}$ is entry-wise nonnegative.
Let $\bm \Delta(t) \coloneq \bm q_2(t)- \bm q_1(t)$ and $\bm q_1(0) = \bm q_2(0)$. Then
\begin{align}
    \bm \Delta(T) =  \int_0^T e^{L_{\mathcal S}(T-s)}L_{\partial \mathcal S}\qty(\bm u_2(s)- \bm u_1(s))\ ds,
\end{align}
holds.
Since $L_{\mathcal S}$ is a Metzler matrix, $e^{L_{\mathcal S}(T-s)}$ is entry-wise nonnegative \cite{farina2011positive}.
Therefore, if $\bm u_1(t)\leq \bm u_2(t)$ holds for all $t\in[0,T]$, then the integrand in the above expression is entry-wise nonnegative.
Hence, $\bm q_1(T)\leq\bm q_2(T)$ holds, and since $\bm p_0\geq \bm 0$, we obtain $y_1(T)\leq\ y_2(T)$.
\qed


\subsection{Proof of Lemma \ref{lemma2}}\label{app.2}
Since $u_{\bm\mu,i}(t)=\mathbb E\qty[\bm X^{\bm \mu}(t)\mid \bm X(0)=\bm x_i]$, if inequality $\qty(\mathcal L\ {\bm x}^{\bm \mu})(\bm x) \leq h_{\bm \mu}^+(\bm x)$ holds, then, by Dynkin's formula \cite{kallenberg1997foundations} under the assumption in Lemma \ref{lemma2}
\begin{align*}
    u_{\bm\mu,i}(t) &\leq \bm x_i^{\bm \mu} + \int_0^t c_{\bm \mu,\bm \mu}^+u_{\bm\mu,i}(s) +\sum_{\substack{|\bm \nu| \leq |\bm \alpha| \\ \bm \nu \neq \bm \mu}} c_{\bm \mu,\bm \nu}^+u_{\bm\nu,i}(s)\ ds .
\end{align*}
Assuming $u_{\bm\nu,i}^-(t) \leq u_{\bm\nu,i}(t)\leq u_{\bm\nu,i}^+(t)$ for $\bm\nu\neq\bm\mu$, we define $u_{\bm\mu,i}^+(t)$ by
\begin{align*}
    u_{\bm \mu,i}^+(t) =&\ \bm x_i^{\bm \mu} + \int_0^t \Big\{c_{\bm \mu,\bm \mu}^+u_{\bm \mu,i}^+(s) \notag \\
    &+\!  \sum_{\substack{|\bm \nu| \leq |\bm \alpha| \\ \bm \nu \neq \bm \mu}}\!\! \qty{\! \qty{c_{\bm \mu,\bm \nu}^+}^+u_{\bm \nu,i}^+(s)\! +\! \qty{c_{\bm \mu,\bm \nu}^+}^-u_{\bm \nu,i}^-(s)}\Big\} ds.
\end{align*}
since the integrand above satisfies 
\begin{align*}
    c_{\bm \mu,\bm \nu}^+u_{\bm\nu,i}(t)\leq \qty{c_{\bm \mu,\bm \nu}^+}^+u_{\bm \nu,i}^+(t)\! +\! \qty{c_{\bm \mu,\bm \nu}^+}^-u_{\bm \nu,i}^-(t),\ \bm\nu\neq \bm\mu. 
\end{align*}
Therefore, by differentiating both sides of the above equation with respect to $t$, we obtain the case for $+$ sign in eq. \eqref{eq:ODE of u^pm}. 
Consequently, if there exists, for every $\bm \mu$ satisfying $|\bm \mu| \leq |\bm \alpha|$, a polynomial $h_{\bm \mu}^+(\bm x)$ that satisfies \eqref{eq:inequality of operator}, then $u_{\bm \alpha,i}^+(t)$ can be obtained by eq. \eqref{eq:ODE of u^pm}.
The case for $-$ sign can be proven in a similar manner by applying Dynkin's formula to the inequality $h_{\bm \mu}^-(\bm x)\leq \qty(\mathcal L\ {\bm x}^{\bm \mu})(\bm x)$.
\qed


\subsection{Proof of Theorem \ref{theorem2}}\label{app.3}
The binomial theorem leads to 
\begin{align}\label{eq:binomial theorem}
    \qty(\mathcal L\ \bm x^{\bm \mu})(\bm x) = \sum_{j=1}^{r}\lambda_j(\bm x)\sum_{\substack{\bm \nu \le \bm \mu\\ |\bm \nu|\ge 1}}\binom{\bm \mu}{\bm \nu}\bm s_j^{\bm \nu}\, \bm x^{\bm \mu-\bm \nu}. 
\end{align}
Since each reaction is elementary, $\lambda_j(x)$ is a polynomial of degree at most two, and the terms of total degree $|\bm\mu|+1$ are
\begin{align}
     \sum_{j\in \mathcal J_{\mathrm{bi}}}\sum_{i=1}^{n} \mu_i s_{j,i}\, \lambda_j(\bm x)\bm x^{\bm \mu-\bm e_i}.
\end{align}
By the assumption $s_{j,i}\leq 0$ holds for all $j \in \mathcal J_{\mathrm{bi}}$ and all $i$, then the above $(|\bm\mu|+1)$-th degree terms are non-positive. 
Dropping these terms from eq. \eqref{eq:binomial theorem} yields
$\qty(\mathcal L\ \bm x^{\bm \mu})(\bm x)\leq h_{\bm \mu}^+(\bm x)$ where $h_{\bm\mu}^+(\bm x)$ is given by \eqref{eq:h+}.
\qed

\begin{thebibliography}{10}
\providecommand{\url}[1]{#1}
\csname url@rmstyle\endcsname
\providecommand{\newblock}{\relax}
\providecommand{\bibinfo}[2]{#2}
\providecommand\BIBentrySTDinterwordspacing{\spaceskip=0pt\relax}
\providecommand\BIBentryALTinterwordstretchfactor{4}
\providecommand\BIBentryALTinterwordspacing{\spaceskip=\fontdimen2\font plus
\BIBentryALTinterwordstretchfactor\fontdimen3\font minus \fontdimen4\font\relax}
\providecommand\BIBforeignlanguage[2]{{%
\expandafter\ifx\csname l@#1\endcsname\relax
\typeout{** WARNING: IEEEtran.bst: No hyphenation pattern has been}%
\typeout{** loaded for the language `#1'. Using the pattern for}%
\typeout{** the default language instead.}%
\else
\language=\csname l@#1\endcsname
\fi
#2}}

\bibitem{khammash2022cybergenetics}
M.~H. Khammash, ``Cybergenetics: Theory and applications of genetic control systems,'' \emph{Proceedings of the IEEE}, vol. 110, no.~5, pp. 631--658, 2022.

\bibitem{briat2023noise}
C.~Briat and M.~Khammash, ``Noise in biomolecular systems: Modeling, analysis, and control implications,'' \emph{Annual Review of Control, Robotics, and Autonomous Systems}, vol.~6, no.~1, pp. 283--311, 2023.

\bibitem{anderson2011continuous}
D.~F. Anderson and T.~G. Kurtz, ``Continuous time {Markov} chain models for chemical reaction networks,'' in \emph{Design and Analysis of Biomolecular Circuits: Engineering Approaches to Systems and Synthetic Biology}.\hskip 1em plus 0.5em minus 0.4em\relax Springer, 2011, pp. 3--42.

\bibitem{gillespie1992rigorous}
D.~T. Gillespie, ``A rigorous derivation of the chemical master equation,'' \emph{Physica A: Statistical Mechanics and its Applications}, vol. 188, no. 1-3, pp. 404--425, 1992.

\bibitem{gillespie2000chemical}
------, ``The chemical {Langevin} equation,'' \emph{The Journal of Chemical Physics}, vol. 113, no.~1, pp. 297--306, 2000.

\bibitem{gillespie1976general}
------, ``A general method for numerically simulating the stochastic time evolution of coupled chemical reactions,'' \emph{Journal of Computational Physics}, vol.~22, no.~4, pp. 403--434, 1976.

\bibitem{munsky2006finite}
B.~Munsky and M.~Khammash, ``The finite state projection algorithm for the solution of the chemical master equation,'' \emph{The Journal of Chemical Physics}, vol. 124, no.~4, 2006.

\bibitem{sukys2022approximating}
A.~Sukys, K.~{\"O}cal, and R.~Grima, ``Approximating solutions of the chemical master equation using neural networks,'' \emph{Iscience}, vol.~25, no.~9, 2022.

\bibitem{naasell2003extension}
I.~N{\aa}sell, ``An extension of the moment closure method,'' \emph{Theoretical Population Biology}, vol.~64, no.~2, pp. 233--239, 2003.

\bibitem{hespanha2008moment}
J.~Hespanha, ``Moment closure for biochemical networks,'' in \emph{2008 3rd International Symposium on Communications, Control and Signal Processing}.\hskip 1em plus 0.5em minus 0.4em\relax IEEE, 2008, pp. 142--147.

\bibitem{singh2010approximate}
A.~Singh and J.~P. Hespanha, ``Approximate moment dynamics for chemically reacting systems,'' \emph{IEEE Transactions on Automatic Control}, vol.~56, no.~2, pp. 414--418, 2010.

\bibitem{naghnaeian2017robust}
M.~Naghnaeian and D.~Del~Vecchio, ``Robust moment closure method for the chemical master equation,'' in \emph{2017 IEEE Conference on Control Technology and Applications (CCTA)}.\hskip 1em plus 0.5em minus 0.4em\relax IEEE, 2017, pp. 967--972.

\bibitem{ghusinga2017exact}
K.~R. Ghusinga, C.~A. Vargas-Garcia, A.~Lamperski, and A.~Singh, ``Exact lower and upper bounds on stationary moments in stochastic biochemical systems,'' \emph{Physical Biology}, vol.~14, no.~4, p. 04LT01, 2017.

\bibitem{sakurai2017convex}
Y.~Sakurai and Y.~Hori, ``A convex approach to steady state moment analysis for stochastic chemical reactions,'' in \emph{2017 IEEE 56th Annual Conference on Decision and Control (CDC)}.\hskip 1em plus 0.5em minus 0.4em\relax IEEE, 2017, pp. 1206--1211.

\bibitem{sakurai2018optimization}
------, ``Optimization-based synthesis of stochastic biocircuits with statistical specifications,'' \emph{Journal of The Royal Society Interface}, vol.~15, no. 138, p. 20170709, 2018.

\bibitem{dowdy2018bounds}
G.~R. Dowdy and P.~I. Barton, ``Bounds on stochastic chemical kinetic systems at steady state,'' \emph{The Journal of Chemical Physics}, vol. 148, no.~8, 2018.

\bibitem{kuntz2019bounding}
J.~Kuntz, P.~Thomas, G.-B. Stan, and M.~Barahona, ``Bounding the stationary distributions of the chemical master equation via mathematical programming,'' \emph{The Journal of Chemical Physics}, vol. 151, no.~3, 2019.

\bibitem{sakurai2022interval}
Y.~Sakurai and Y.~Hori, ``Interval analysis of worst-case stationary moments for stochastic chemical reactions with uncertain parameters,'' \emph{Automatica}, vol. 146, p. 110647, 2022.

\bibitem{sakurai2018bounding}
------, ``Bounding transient moments of stochastic chemical reactions,'' \emph{IEEE Control Systems Letters}, vol.~3, no.~2, pp. 290--295, 2018.

\bibitem{dowdy2018dynamic}
G.~R. Dowdy and P.~I. Barton, ``Dynamic bounds on stochastic chemical kinetic systems using semidefinite programming,'' \emph{The Journal of Chemical Physics}, vol. 149, no.~7, 2018.

\bibitem{holtorf2024tighter}
F.~Holtorf and P.~I. Barton, ``Tighter bounds on transient moments of stochastic chemical systems,'' \emph{Journal of Optimization Theory and Applications}, vol. 200, no.~1, pp. 104--149, 2024.

\bibitem{norris1998markov}
J.~R. Norris, \emph{Markov chains}.\hskip 1em plus 0.5em minus 0.4em\relax Cambridge university press, 1998, no.~2.

\bibitem{briat2016antithetic}
C.~Briat, A.~Gupta, and M.~Khammash, ``Antithetic integral feedback ensures robust perfect adaptation in noisy biomolecular networks,'' \emph{Cell systems}, vol.~2, no.~1, pp. 15--26, 2016.

\bibitem{gardner2000construction}
T.~S. Gardner, C.~R. Cantor, and J.~J. Collins, ``Construction of a genetic toggle switch in {Escherichia} coli,'' \emph{Nature}, vol. 403, no. 6767, pp. 339--342, 2000.

\bibitem{farina2011positive}
L.~Farina and S.~Rinaldi, \emph{Positive linear systems: theory and applications}.\hskip 1em plus 0.5em minus 0.4em\relax John Wiley \& Sons, 2011.

\bibitem{kallenberg1997foundations}
O.~Kallenberg, \emph{Foundations of modern probability}.\hskip 1em plus 0.5em minus 0.4em\relax Springer, 1997.

\end{thebibliography}
\end{document}